# Assessment of the proton boron fusion reaction for practical radiation therapy applications using MCNP6

David P. Adam[1] and Bryan Bednarz[2]

[1] Department of Engineering Physics, University of Wisconsin-Madison, 1500 Engineering Dr. Madison, WI 53706

[2] Department of Medical Physics, University of Wisconsin-Madison, 1111 Highland Ave., Madison, WI 53706

(Dated: 28 July 2015)

The proton boron fusion reaction, $^{11}$B(p, 3α), is investigated using MCNP6 to assess the viability for potential use in radiation therapy. The $^{11}$B(p, 3α) is theoretically desirable because it could combine the localized dose delivery protons exhibit (Bragg peak) with the creation of three high LET α particles that locally deposit energy in tumor regions. Simple simulations of a proton pencil beam incident upon a water phantom and a water phantom with an axial region also containing boron were modeled using MCNP6 in order to determine the extent of the impact boron had upon the calculated energy deposition. The MCNP simulations performed demonstrated that the proton boron fusion reaction rate at clinically relevant boron concentrations was too small in order to have any measurable impact on the absorbed dose. It was determined that there are no good evaluations of the $^{11}$B(p, 3α) reaction for use in MCNPX/6 and further work should be conducted in cross section evaluations in order to definitively evaluate the feasibility of the proton boron fusion reaction for use in radiation therapy applications.

Keywords: Proton boron fusion reaction, MCNP6, Proton therapy

The proton-boron fusion reaction is an aneutronic fusion reaction that has been studied extensively for commercial fusion power applications[1]. The reaction describes the creation of three alpha particles as the result of the interaction of a proton incident upon a $^{11}$B target. The $^{11}$B target nucleus is transmuted into an excited $^{12}$C, which then decays by emission of an 3.76 MeV alpha particle into an excited $^{8}$Be. In turn, the excited $^{8}$Be nucleus subsequently decays into two 2.74 MeV alpha particles[1]. Theoretically, the proton boron fusion reaction would be a desirable reaction to exploit during radiation therapy in that, with the appropriate boron delivery agent, it could potentially combine the localized dose delivery of protons beams (Bragg peak) with the local deposition and high LET of alpha particles in cancerous sites.

The potential use of the proton-boron reaction in the context of proton radiation therapy was originally proposed by Yoon et al[2]. The group demonstrated that the proton-boron fusion reaction dramatically enhanced the maximum absorbed dose in a simulated tumor region and could potentially improve the efficacy of proton therapy. In an effort to replicate the MCNP results of proton boron-fusion reaction simulations presented by Yoon et al[2], we discovered that the dosimetric enhancement reported by these authors is greatly overestimated. The overestimation is likely due to the inappropriate use of default reaction cross sections in MCNPX for their simulation. For this investigation a detailed evaluation of cross sections for protons incident upon an $^{11}$B target available for use in MCNP was performed and a more accurate set of dose enhancement factors were determined.

The default LA150h proton cross section library that comes with MCNPX v2.7.0 does not include cross sections for protons incident upon $^{11}$B. Cross section evaluations for protons incident upon $^{11}$B can be found in the TENDL library[3], which is evaluated using the TALYS code package[4] and are generated each year and designated according to the year they were generated (i.e. TENDL-2009, TENDL-2010, TENDL-2011, etc.). There are multiple issues with the TENDL cross section evaluations. They adhere to ENDF-6 formatting standards; as such, there are standard reaction types designated by specific "MT" numbers. The only "MT" number for which the TENDL cross sections are defined for in the available *.ace files used by MCNPX/6 is the "MT=5" reaction, which is a general nuclear heating reaction type and does not simulate the transport of any specific particles or reaction types. Of secondary concern, there is very little experimental data for protons incident upon $^{11}$B to corroborate the results of the TALYS code and as such, the evaluated cross sections cannot be fully trusted to be representative of these nuclear processes. There has been some work in experimentally evaluating cross sections for protons incident upon $^{11}$B, but almost all work has been done near or below 1 MeV incident protons and this data is unavailable for use in MCNPX/6[4-6].

Given that cross section evaluations in MCNPX/6 for protons incident on $^{11}$B were not available, an alternative evaluation method was required in order to simulate the effect of boron in the water phantom. A report by Soppera et al[7] was the only citation found by the authors where any cross section data for the applicable reaction types to protons incident on $^{11}$B target. The report gives cross section data in charts from the TENDL-2011 evaluated nuclear library for

$^{11}$B(p,α)$^8$Be (MT=107) and $^{11}$B(p,2α)$^4$He (MT=108) reactions. Although the uncertainty in the cross sections could be substantial, experimental data of the $^{11}$B(p,3α) reported for low energies is consistent with the magnitude of the evaluated cross sections[4-6]. No evaluated data could be found for $^{11}$B(p,3α) (MT=109) so for this work it was assumed that the summation of the $^{11}$B(p,α)$^8$Be (MT=107) and $^{11}$B(p,2α)$^4$He reaction rates will suffice to simulate the $^{11}$B(p,3α) reaction.

In order to simulate the scenario where boron is present in the medium, the DE/DF dose response cards were used as flux tally modifiers to calculate reaction rates for both the $^{11}$B(p, α)$^8$Be and $^{11}$B(p, 2α)$^4$He reactions. The TENDL-2011 boron cross section data used for the DE/DF cards was extracted from Soppera et al[7] using the data extraction tool WebPlotDigitizer[8] and the cross sections used in the simulation are presented in Figure 1. This methodology may not be valid for high concentrations of boron, but should be valid for typical clinical applications where boron concentrations are on the order of 10-100 ppm[9]. It should be noted that these cross sections are on the order of hundreds of millibarns and the presence of small concentrations of boron should not perturb the flux dramatically.

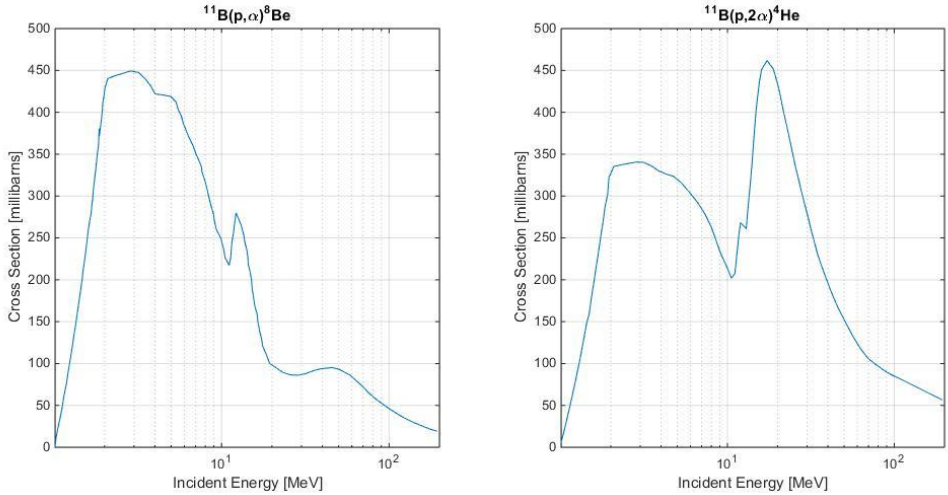

**FIG. 1:** TENDL-2011 Cross section evaluations for $^{11}$B reactions

A cylindrical water phantom was modeled with a 8 cm radius and various depths reflecting the range of different proton energies considered. The materials used were hydrogen (ZAID=1001.70h) and oxygen (ZAID=8016.70h) from the LA150h library with an atomic ratio of 2:1 respectively. The assigned mass density of water was 1 g/cm$^3$. A baseline case was established in which the energy deposition for all particles was tabulated for only the water phantom without the presence of boron. The second case used the DE/DF cards for an applicable axial slice of the phantom that coincided with the Bragg peak of the incident protons. Since the DE/DF cards only modify the flux of the protons in water and the validity of the cross sections is suspect, the axial slice is not necessarily representative of the actual dose imparted to the volume, it is merely meant to demonstrate the potential dose response within the illustrated region. This axial slice will be referred to as the Boron Uptake Region (BUR)[2]. For example, the BUR occurred around 4 to 6 cm deep into the phantom for an 80 MeV incident proton beam.

A mono-energetic proton pencil beam source was defined 50 cm away from the end face of the cylinder; the incident beam was centered and perpendicular to the flat face of the cylinder. All protons travel axially through the cylinder. Four different initial proton beam energy scenarios were: 80 MeV, 90 MeV, 100 MeV, and 250 MeV. A total of $1 \times 10^6$ million particle histories were ran for each simulation.

The two primary tallies used to quantify the impact of boron within the BUR are surface flux tallies (TMESH type 1, the equivalent of an f4 tally), and energy deposition tallies (TMESH type 3, the equivalent of an +f6 tally). To quantify the absorbed dose of the water phantom in the baseline case, the TMESH type 3 mesh tally was used. The TMESH type 3 is technically a tally that quantifies the energy transferred to the medium by all particles, thus technically making it a collisional kerma tally and not an absorbed dose tally. However, for this calculation, it is allowable to assume that charged particle equilibrium exists within the phantom making the calculated collisional kerma is equivalent to the absorbed dose[10].

Cross section data for hydrogen and oxygen was defined in the materials input deck for the BUR. The DE/DF dose response cards were used to calculate the reaction rate within the BUR for both $^{11}B(p, \alpha)^8Be$ and $^{11}B(p, 2\alpha)^4He$ reactions of the TMESH type 1 flux tally. The reaction rates for both reactions were both considered separately and were summed in the results. The absorbed dose was calculated according to a mesh tally multiplier (FM4) in Equation 1 by assuming that α particles have a quality factor (Q) of 20, the concentration of the boron is 100 ppm (c), the energy of three α particles released from the $^{11}B(p, 3\alpha)$ reaction is equivalent to 8.68 MeV[1] and is deposited locally,

$$FM4 = Q \times E \times c = 20 \times 8.68 \, \text{MeV} \times 100 \, \text{ppm} = 0.01736 \qquad (1)$$

Note that it was assumed that all energy deposition from the proton-boron reaction remains at the site of interaction, which is a reasonable assumption considering the range of α particles in water at the energies considered.

Figure 2 presents the calculated relative absorbed dose for the 80 MeV proton beam where the BUR contains a boron concentration of 100 ppm using the TENDL-2011 cross sections. All calculated results are normalized to the maximum value of the dose computed for the pure water phantom. As depicted in the figure, the maximum relative dose of the summed $^{11}$B(p, α)$^8$Be and $^{11}$B(p, 2α)$^4$He reactions is 0.019% of the maximum dose. It must be re-emphasized that MCNP simulations are only as accurate as the cross sections defined and in this simulation the cross sections are not validated by experimental data. Thus, the results should be taken as an indicator of whether or not the calculated absorbed dose of the $^{11}$B(p, 3α) reaction is within the same order of magnitude as the calculated absorbed dose for the pure water phantom.

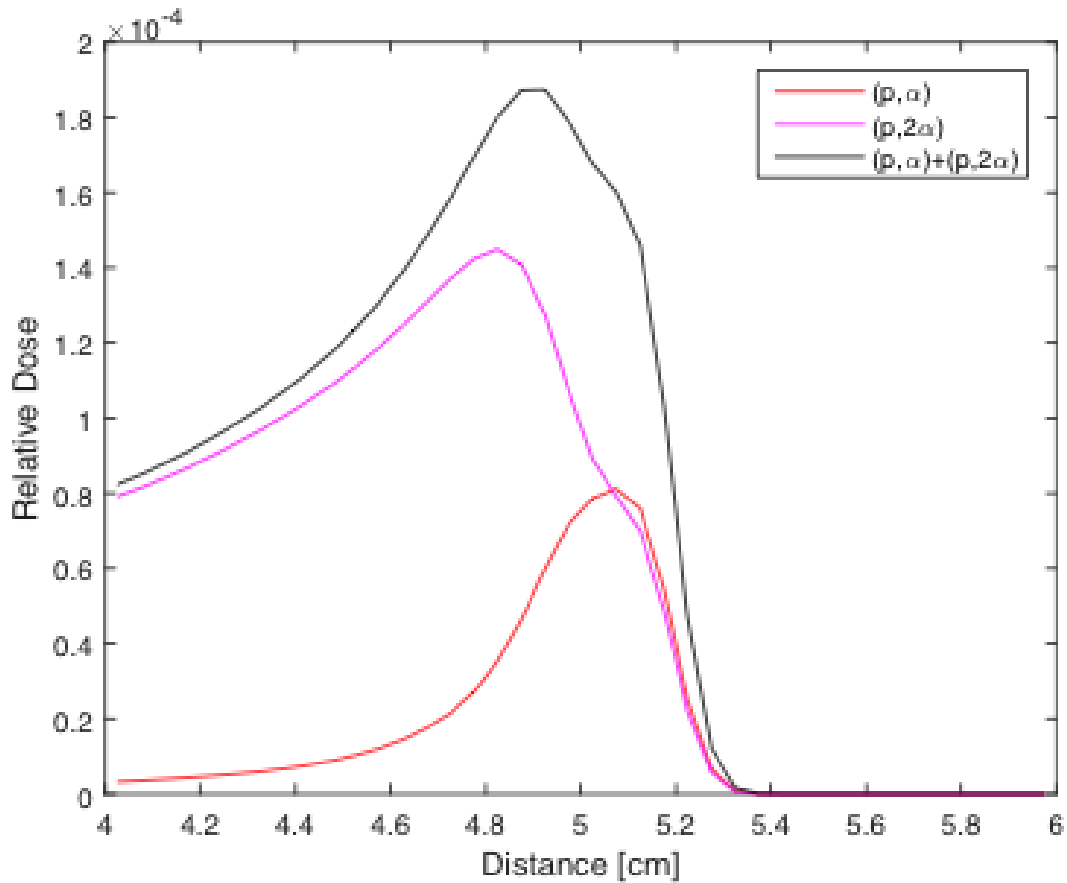

**FIG. 2:** Relative dose for $^{11}$B reaction types assuming 100 ppm boron to water concentration using TENDL-2011 cross sections.

Figure 3 presents the calculated relative absorbed dose in the phantoms for beam energies of 80 MeV (Figure 3a), 90 MeV (Figure 3b), 100 MeV (Figure 3c), and 250 MeV (Figure3d). The region in the phantom for which the reaction rates were calculated are isolated by the vertical lines in the figures. Given the large differences between the absorbed dose in the BUR to the absorbed dose in the pure water phantom, the relative dose of the pure water phantom was scaled by a factor of 1/10000. The horizontal line is given as a reference for a 1/10000 of the maximum dose in water. For the BUR region, the relative dose in the BUR increases as the beam energy is increased, but not significantly. The maximum dose calculated was for the 250 MeV beam, and the summed $^{11}$B(p, α)$^8$Be and $^{11}$B(p, 2α)$^4$He dose was only 0.026% of the maximum computed dose.

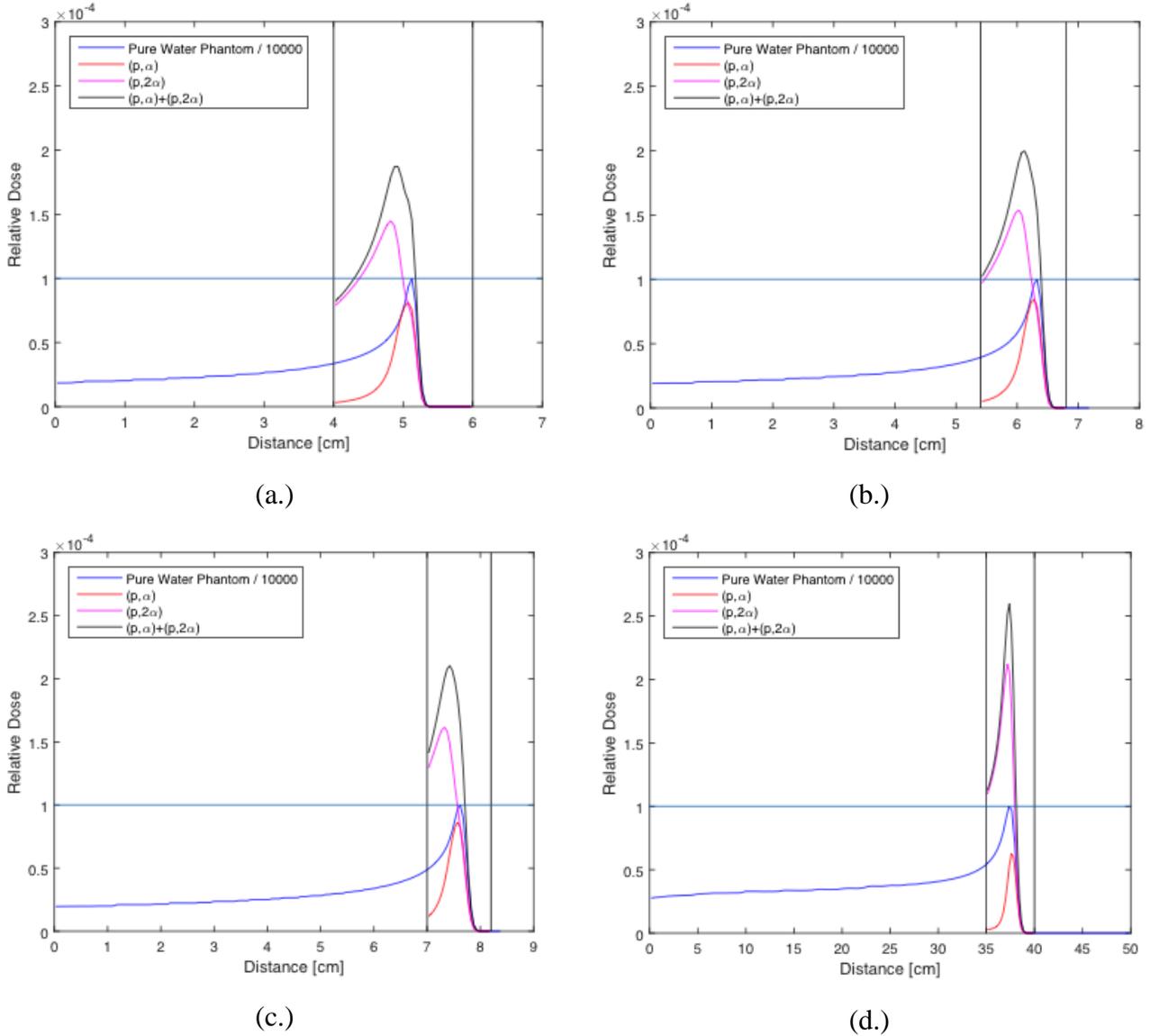

(a.)  (b.)

(c.)  (d.)

**FIG. 3:** Relative Dose for Pure Water Phantom and $^{11}$B Reaction Types with Water Phantom Dose Divided by 10,000 using TENDL-2011 Cross Sections

For all MCNP6 simulations conducted, the increase of absorbed dose of a simple water phantom due to the $^{11}$B(p, 3α) reaction was found to be inconsequential. Using the most appropriate available cross section data, we show that the calculated dose enhancement reported by Yoon et al[2] is likely to be significantly overestimated. The calculated increased absorbed dose due to the $^{11}$B(p, 3α) reaction calculated from this work and the work performed by Yoon et al[2] are reported in Table I.

TABLE I: Comparison of overall increase of maximum dose for each reaction type and comparison to Yoon et al[2].

| Incident Energy [MeV] | (p, α) | (p, 2α) | (p, α)+(p, 2α) | Yoon et al |
|---|---|---|---|---|
| 80 | 0.0081 | 0.0145% | 0.0187% | 50.5% |
| 90 | 0.0084 | 0.0154% | 0.0200% | 50.9% |
| 100 | 0.0086 | 0.0162% | 0.0210% | 79.5% |
| 250 | 0.0063 | 0.0212% | 0.0260% | N/A |

The most problematic aspect of the simulation was the lack of available cross section data for any reaction types specific to protons incident upon a 11B for use in MC- NPX/6. Therefore, an approximation was made by assigning the cross section data found for the $^{11}$B(p, α)$^8$Be and the $^{11}$B(p, 2α)$^4$He reaction types to the DE/DF cards in MCNP6 to calculate the reaction rates of each reaction type. Most importantly, in order to accurately model the problem, cross section data for protons incident upon $^{11}$B must be properly evaluated. The results of this simulation are meant as a sanity check to evaluate the realistic prospects of using the proton boron fusion reaction in clinical radiotherapy applications.

Special thanks to Professor Paul Wilson and Dr. Tim Bohm for their assistance with this work.

**Word Count: 3451**

[1] D. Moreau, "Potentiality of the proton-boron fuel for controlled thermonuclear fusion," Nuclear Fusion 17, 13 (1977).
[2] D.-K. Yoon, J.-Y. Jung, and T. S. Suh, "Application of proton boron fusion reaction to radiation therapy: A Monte Carlo simulation study," Applied Physics Letters 105, 223507 (2014).
[3] S. van der Marck, A. Koning, and D. Rochman, "Benchmarking tendl-2012," Nuclear Data Sheets 118, 446 – 449 (2014).

[4]A. Koning and D. Rochman, "Modern nuclear data evaluation with the {TALYS} code system," Nuclear Data Sheets 113, 2841– 2934 (2012), special Issue on Nuclear Reaction Data.

[5]J. Davidson, H. Berg, M. Lowry, M. Dwarakanath, A. Sierk, and P. Batay-Csorba, "Low energy cross sections for $^{11}$B(p, 3alpha)," Nuclear Physics A 315, 253 – 268 (1979).

[6]L. Lamia, S. Puglia, C. Spitaleri, S. Romano, M. G. D. Santo, N. Carlin, M. G. Munhoz, S. Cherubini, G. Kiss, V. Kroha, S. Kubono, M. L. Cognata, C.-B. Li, R. Pizzone, Q.-G. Wen, M. Sergi, A. S. de Toledo, Y. Wakabayashi, H. Yam-aguchi, and S.-H. Zhou, "Indirect study of $^{11}$B(p,alpha)$^{8}$Be and $^{10}$B(p,alpha)$^{7}$Be reactions at astrophysical energies by means of the trojan horse method: recent results," Nuclear Physics A 834, 655c – 657c (2010), the 10th International Conference on Nucleus-Nucleus Collisions (NN2009).

[7]M. B. N. Soppera, E. Dupont, "Janis book of proton-induced cross sections," (2012).

[8]A. Rohatgi, "Webplotdigitizer," (2010).

[9]W. F. Verbakel, W. Sauerwein, K. Hideghety, and F. Stecher- Rasmussen, "Boron concentrations in brain during boron neutron capture therapy: in vivo measurements from the phase I trial {EORTC} 11961 using a gamma-ray telescope," International Journal of Radiation Oncology*Biology*Physics 55, 743 – 756 (2003).

[10]S. B. Jia, M. H. Hadizadeh, A. A. Mowlavi, and M. E. Loushab, "Evaluation of energy deposition and secondary particle produc- tion in proton therapy of brain using a slab head phantom," Reports of Practical Oncology & Radiotherapy 19, 376 – 384 (2014).